\documentclass[letterpaper]{jpconf}
\usepackage{graphicx}
\begin{document}
\title{Probing Dense Partonic Matter at RHIC}

\author{Barbara Jacak for the PHENIX Collaboration}

\address{Department of Physics and Astronomy \\
Stony Brook University,\\
Stony Brook, NY 11794, USA}

\ead{Barbara.Jacak@stonybrook.edu}

%\begin{abstract}
%\end{abstract}.

\section{Introduction}

Comprehensive analysis of the experimental data collected by the
PHENIX experiment at the Relativistic Heavy Ion Collider (RHIC)
during the first three years of operation has shown that a new
state of matter is created in Au+Au collisions. This matter cannot
be described in terms of ordinary color neutral hadrons \cite{phenix_wp}.
Strongly interacting systems under extreme conditions of temperature
and density have long been predicted to undergo a phase transition 
where their constituent quarks and gluons are no longer confined 
into hadrons.

To characterize the properties of this new matter, it is instructive
to review approaches to characterize electromagnetic plasmas. Plasma
physicists typically seek to determine the pressure, viscosity and
equation of state, along with the thermalization time and extent of 
plasmas as they become experimentally accessible. It should be noted
that not all electromagnetic plasmas are very long lived; the lifetime
of plasmas created by high intensity laser light impinging upon a
thin target is only a few nanoseconds. 

These same quantities are of interest for the quark gluon plasma. 
They can be
determined from measurements of collective behavior among particles 
emitted from the plasma or by probing with 
energetic particles 
transmitted through the plasma. Other plasma properties of interest for
both electromagnetic and strong interaction plasmas are radiation
rate, collision frequency, conductivity, opacity/transmision probability, 
and the Debye screening length. These are best determined using
probe particles with De Broglie wavelengths short compared to the characteristic
wavelength of the plasma. For electromagnetic plasmas, the probes
of choice are energetic photons or electrons impinging upon the plasma
from the outside. The analogous probes of the quark gluon plasma
are high momentum quarks and gluons produced in the very
first nucleon-nucleon collisions before the processes driving
the system toward thermalization take place. In these proceedings,
I will focus primarily on new measurements of such probes.

The PHENIX experiment is optimized to measure electromagnetic probes
and high transverse momentum phenomena, along with global variables,
multiparticle flows, and soft identified hadron spectra to understand
the evolution of the produced matter over all relevant timescales.
These diverse criteria required combining a large number of detection
subsystems with a high bandwidth trigger and data acquisition system.
A description of the PHENIX spectrometers can be found in reference
\cite{phenix_nim}.

\section {Initial State}

In order to search for effects of the matter upon the probes, it
is essential to first benchmark the production of those probes in
p+p collisions, and determine the effects of binding in the
intial state nucleus upon the distribution of partons in the 
colliding nucleons. PHENIX measured the yields as a function
of transverse momentum of $\pi^0$\cite{pp_pi0} 
and direct photons\cite{pp_photons} in $\sqrt{s}$ = 200 GeV
p+p collisions. Both the yields and the $p_T$ distributions are 
well reproduced by leading order pQCD\cite{pp_pi0,pp_photons}..

%%%%%%%%%%%%%%%%%%%%%%%%%%%%%%%%%%%%%%%%%%%% Fig. 1
\begin{figure}
\begin{center}
\includegraphics[width=0.6\linewidth]{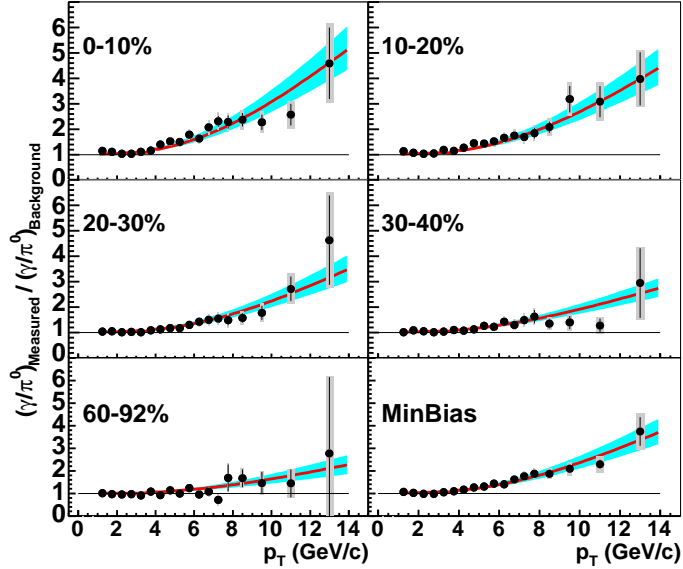}
\caption{\label{fig:photons} 
Double ratio of measured to background $(\gamma/\pi^0)$
as a function of $p_T$ for minimum bias and five centralities of 
Au+Au collisions at $\sqrt{s_{NN}}$ = 200 GeV \cite{auauphotons}.
Solid curves are the ratio of pQCD predictions to the background
photon yield based on the measured $\pi^0$ for each centrality
class. The shaded regions around the curves indicate variation of the pQCD
calculation for scale changes from $p_T/2$ to $2\ p_T$ plus the
$\langle N_{coll}\rangle$ uncertainty.
}
\end{center}
\end{figure}
%%%%%%%%%%%%%%%%%%%%%

As photons are not expected to interact with the color charges
of quarks and gluons in the system, direct photon production 
in Au+Au collisions should also agree with calculations
based upon pQCD. In PHENIX this expectation is tested in
$\sqrt{s_{NN}}$ = 200 GeV Au+Au collisions by
measuring the ratio of all photons to those from $\pi^0$ decays,
and plotting the double ratio of measured to simulated
$\gamma/\pi^0$. As the simulation is performed starting with
the measured $\pi^0$ spectrum, the $\pi^0$ contribution cancels in the
double ratio, making it the ratio of inclusive to direct photons.
This technique minimizes systematic uncertainties.
Any significant deviation of the double ratio above unity
indicates a direct photon excess \cite{Stefan}. 
Figure~\ref{fig:photons} shows
this double ratio as a function of $p_T$ in Au+Au collisions
in different centrality bins \cite{auauphotons}.
An excess is observed at high $p_T$
with a magnitude that increases with increasing centrality of 
the collision. The measured results are compared to same NLO
pQCD predictions which agreed with the PHENIX p+p data \cite{NLO}, 
scaled by the number of binary
nucleon collisions for each centrality class. The binary scaled
predictions agree well with the measured direct
photons\cite{auauphotons}. 
The increasing ratio with centrality is attributed to
the decreasing decay background due to $\pi^0$ suppression \cite{ppg014}.

The agreement of the direct photon production rate in Au+Au collisions
with NLO pQCD calculations indicates that the effects of nucleons being
bound inside a nucleus do not preclude the use of perturbative QCD.
This is remarkable, given the complicated nuclei involved. PHENIX
has also measured the production of charmed mesons in Au+Au and p+p
collisions. We find that the total cross section of charmed mesons 
also scales with the number binary nucleon collisions \cite{charm}. 
As the charm mass is large, $c-\overline{c}$ pairs can only be produced
in high momentum transfer collisions, and should therefore be
calculable with pQCD. The photon results suggest that binary collision
scaling should hold for charm production as well, and PHENIX has
established that it does.
 
%\subsection{d+Au Collisions}

Several initial state effects may be expected in collisions involving
nuclei. These include nuclear shadowing, saturation of the gluon 
distribution, and multiple initial state scattering of the incoming
partons. The collision scaling of high $p_T$ photons and charm yields
indicate that these effects mostly cancel for large $q^2$ processes
at mid-rapidity. However, they should be more important for softer
processes and at rapidities nearer those of the beams. 
A wide range of d+Au collision data was collected to quantify
these effects.  PHENIX can
probe saturation effects via the rapidity dependence of hadron production,
and initial state multiple scattering via the dependence of hadrons
upon the number of collisions suffered by each deuteron participant.
We find that calculations incorporating shadowing and initial state
semi-hard scattering can reproduce the data on hadron production
at midrapidity reasonably well \cite{accardi}. 

%%%%%%%%%%%%%%%%%%%%%%%%%%%%%%%%%%%%%%%%%%%% Fig. 2
\begin{figure}
\begin{center}
\includegraphics[width=0.6\linewidth]{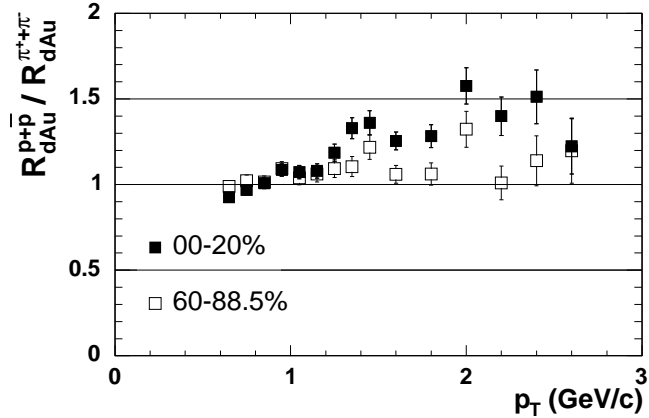}
\caption{\label{fig:ratiordau} 
Ratio of the proton over the pion nuclear modification in d+Au
collisions for central and peripheral events. Error bars indicate
statistical errors only.
}
\end{center}
\end{figure}
%%%%%%%%%%%%%%%%%%%%%
 
However, no model of initial state multiple scatterings can predict
the difference in baryon and meson yields in nucleon- or deuteron-
nucleus collisions. Baryon yields at moderate $p_T$ are considerably
enhanced in central d+Au collisions compared to p+p.  This contrasts
with the very small enhancement observed for pions, and is illustrated
in Figure~\ref{fig:ratiordau}. The figure shows the ratio of R$_{dAu}$
for baryons to that of mesons in the most peripheral (60-88.5\%) and
most central (0-20\%) d+Au collisions. 
The nuclear modification factor, $R_{dAu}$ is defined as
$$ R_{dAu} = \frac{(1/N^{evt}_{dAu}) d^2N_{dAu}/dy dp_T} {T_{dAu}
d^2\sigma^{pp}_{inel}/dy dp_T},  $$
where $T_{dAu} = \langle N_{coll}\rangle/\sigma^{pp}_{inel}$ describes
the nuclear geometry and $d^2\sigma^{pp}_{inel}/dy dp_T$ for p+p
collisions is derived from the measured p+p cross section. 
$\langle N_{coll}\rangle$ is the average number of inelastic nucleon-
nucleon collisions determined from a Glauber simulation.
The figure shows that the difference between baryons
and mesons reaches 50\% in central collisions.
More successful explanations of the baryon enhancement invoke
hadronization by recombination of quarks from fragmenting jets
with those drawn from the nearby nuclear medium \cite{Hwa}.

%%%%%%%%%%%%%%%%%%%%%%%%%%%%%%%%%%%%%%%%%%%% Fig. 3
\begin{figure}
\begin{center}
\includegraphics[width=0.7\linewidth]{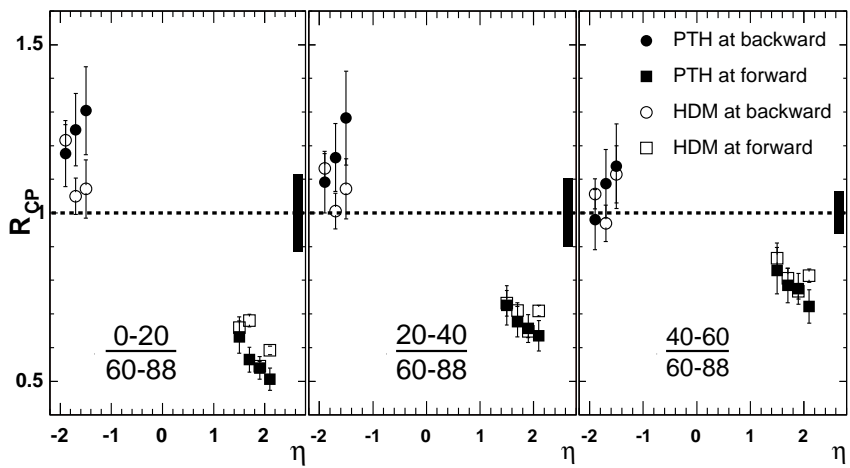}
\caption{\label{fig:rcp} 
$R_{cp}$ as a function of $\eta$ for 1.5$\langle p_T \rangle$
4.0 GeV/c for different centrality classes.
}
\end{center}
\end{figure}
%%%%%%%%%%%%%%%%%%%%%

The rapidity dependence of charged hadron production in d+Au collisions
is measured by PHENIX using the muon arms, which cover
$\eta$ = 1.4 - 2.2. We detect stopped muons
from hadron decays and also hadrons which punch through the absorber
and interact in the muon arm \cite{Anuj}. As comparable data from 
p+p collisions are as yet unavailable, we calculate a nuclear modification
factor by comparing yields in central and peripheral collisions, 
scaled by the corresponding number of binary nucleon collisions
in each centrality bin,~$\langle N_{coll}\rangle$:
$$
  R_{cp} = \frac{ \langle \left( \frac {dN}{d\eta dp_T} \right)^{Central} \rangle / \langle N_{coll}^{Central} \rangle }  {\langle \left(\frac{dN}{d\eta dp_T}\right)^{Peripheral} \rangle / \langle N_{coll}^{Peripheral} \rangle }
$$
%$$ R_{cp} = \frac{\langle(dN/d\eta dp_T)^{central}\rangle / 
%\langle N_{coll}^{central}\rangle} 
%{\frac{\langle(dN/d\eta dp_T)^{peripheral}\rangle / 
%\langle N_{coll}^{peripheral}\rangle} $$
$R_{cp}$ integrated in the $p_T$ range from 1.5 to 4.0 GeV/c
is shown in Figure~\ref{fig:rcp} as a function of $\eta$ for the
different centrality classes \cite{hadronrcp}.
We observe that $R_{cp}$ shows a suppression at forward rapidity
(deuteron going direction)
that is largest for the most central events. The opposite trend
is observed at backward rapidity (Au going direction),
where $R_{cp}$ shows an enhancement
that is also largest for the most central events. There is a weak
$p_T$ dependence, with slightly smaller $R_{cp}$ at lower $p_T$.
We observe a clear pseudorapidity dependence at forward rapidity 
with $R_{cp}$ dropping further at larger $\eta$ values; our
results are consistent with the BRAHMS data \cite{brahms_rcp}. Within
the uncertainties, we are unable to discern any pseudorapidity 
dependence at the backward rapidity.
Forward suppression is qualitatively consistent with several theories
including shadowing or saturation effects in initial state multiple 
scattering, but also recombination \cite{dima,ramona,hwa2}. 
The enhancement at backward rapidity has not yet been explained. 

\section{Thermalization}

Turning now from the initial state, I discuss observables that
tell about the approach to thermalization in Au+Au collisions.
PHENIX measures many species of hadrons, including the $\phi$
meson via its decays to $K^+K^-$ and $e^+e^-$. We find that
the slope of the $\phi~ p_T$ distributions reconstructed in the
two decay channels are consistent with each other, and are also
consistent with a blast wave fit to $\pi, K,$ and p spectra \cite{ppg016}.
This observation supports the picture of a common expanding source
for all low and intermediate momentum hadrons at midrapidity.

%\subsection{Collective Flow}

Good tests of the system's approach toward thermalization
are the magnitude and species dependence of the elliptic flow.
Elliptic flow is measured by Fourier analysis of the momentum
distribution of emitted particles as a function of the 
azimuthal angle with respect to the plane of the Au+Au
collision. The strength of the flow is quantified by the
magnitude of the second harmonic coefficient of the Fourier
expansion, generally known
as the $v_2$ parameter. PHENIX has measured $v_2$
for charged pions, kaons, protons, antiprotons, neutral pions,
inclusive photons and electrons from the decay of charmed mesons.
This is measured using several methods, including two
particle correlations at midrapidity, two-particle cumulants,
and the distribution of particles in the central arm with 
respect to the reaction plane determined by the beam-beam counters
(BBC) at rapidity $\eta$ = 3.0 - 3.9. Most of the measurements 
for identified particles use the PHENIX central arm spectrometers
and the BBC reaction plane. We find that the different
species exhibit approximately the same $v_2$ per 
constituent quark. $v_2$ rises
as a function of $p_T$ up to 1.5-2 GeV/c per constituent quark,
then falls again at higher $p_T$. In $\sqrt{s_{NN}}$ = 62.4 GeV 
Au+Au collisions, the observed $v_2$ per quark is quite similar
to that in 200 GeV Au+Au collisions. 

%%%%%%%%%%%%%%%%%%%%%%%%%%%%%%%%%%%%%%%%%%%% Fig. 4
\begin{figure}
\begin{center}
\includegraphics[width=0.6\linewidth]{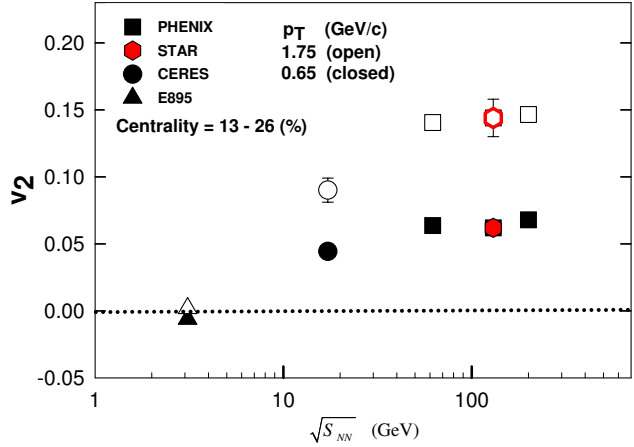}
\caption{\label{fig:v2} 
Differential $v_2$ vs. $\sqrt{s_{NN}}$ for charged hadrons in
nucleus-nucleus collisions. Results are shows for a centrality
class of 13-26\% and $p_T$ of 1.75 GeV/c (open symbols) and 0.65
GeV/c (closed symbols). 
}
\end{center}
\end{figure}
%%%%%%%%%%%%%%%%%%%%%

This is illustrated in
Figure~\ref{fig:v2}, which shows $v_2$ for inclusive charged 
hadrons at two $p_T$ values (0.65 and 1.75 GeV/c), 
as a function of $\sqrt{s}$ for
heavy ion collisions from AGS through RHIC energies \cite{ppg047}.
The data are compared for relatively central collisions representing
the upper 13-26\% of the total cross section. For both $p_T$
cuts, the magnitude of $v_2$ shows a significant increase with
collision energy ($\approx$ 50 \% increase from SPS to RHIC)
up to $\sqrt{s_{NN}}$ = 62.4 GeV. Thereafter, $v_2$ 
appears to saturate for larger beam energies. Given the fact
that the energy density is estimated to increase by approximately
30\% over the range $\sqrt{s_{NN}}$ = 62.4 - 200 GeV, this
apparent saturation of $v_2$ may be indicative of the role of a
rather soft equation of state. Such a softening could result from
the production of mixed phase within this energy range.

%\subsection{Hydrodynamic Models}

Of course, for such a discussion to be applicable, the system must
be in local thermal equilibrium. For thermalization to occur, the
particles must interact and/or radiate. If such processes occur
with sufficient frequency, the system
should be describable by locally equilibrated fluid elements, with
a velocity profile following $\beta_\parallel^{Fluid} = z/t$. This
has been tested by calculations with hydrodynamical models of Au+Au
collisions with initial conditions fixed by the collision geometry
and energy density fixed by total particle and energy production;
for more detail
see discussion and references in \cite{phenix_wp}. In a hydrodynamic
picture, the source of elliptic flow is the spatial anisotropy of the
energy density in the transverse plane at the time hydrodynamics
becomes valid (i.e. at the time that local thermal equilibrium is
reached). If local equilibration and the onset of hydrodynamical
behavior is delayed because interactions between the initially
produced particles are weak at first, then the spatial anisotroy
giving rise to elliptic flow is reduced. Consequently, the
magnitude of the $v_2$ parameter is sensitive to the thermalization
time. The thermalization time may 
be inferred from the hydrodynamics calculations
constrained with data on the $p_T$ and species dependence of
elliptic flow and particle spectra. 

%%%%%%%%%%%%%%%%%%%%%%%%%%%%%%%%%%%%%%%%%%%% Fig. 5
\begin{figure}
\begin{center}
$\begin{array}{c@{\hspace{0.01cm}}c}
%	\multicolumn{1}{1}{\mbox{\bf (a)}} &
%	\multicolumn{1}{1}{\mbox{\bf (b)}} \\ [-0.53cm]
   \includegraphics[width=0.51\linewidth]{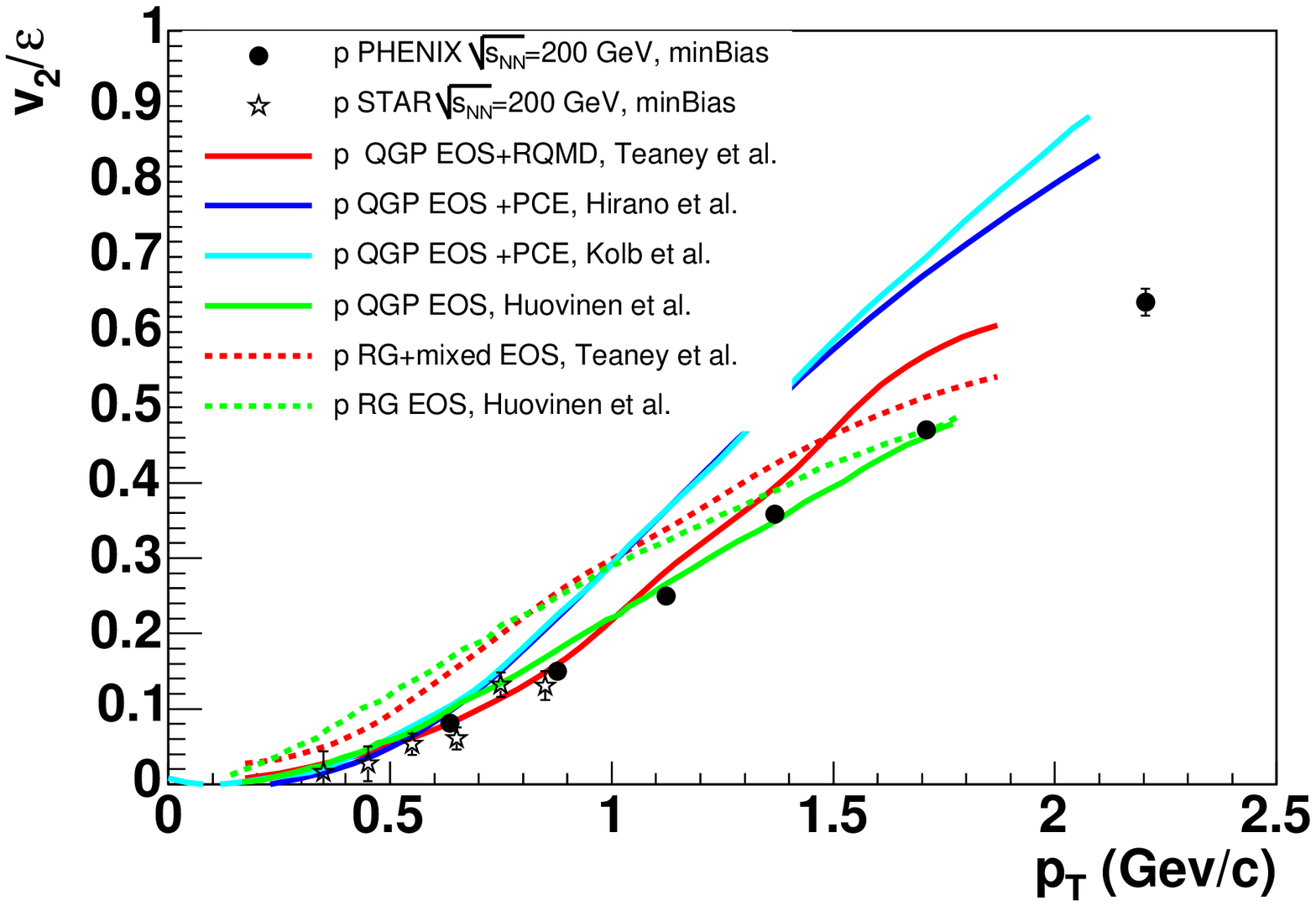} &
   \includegraphics[width=0.51\linewidth]{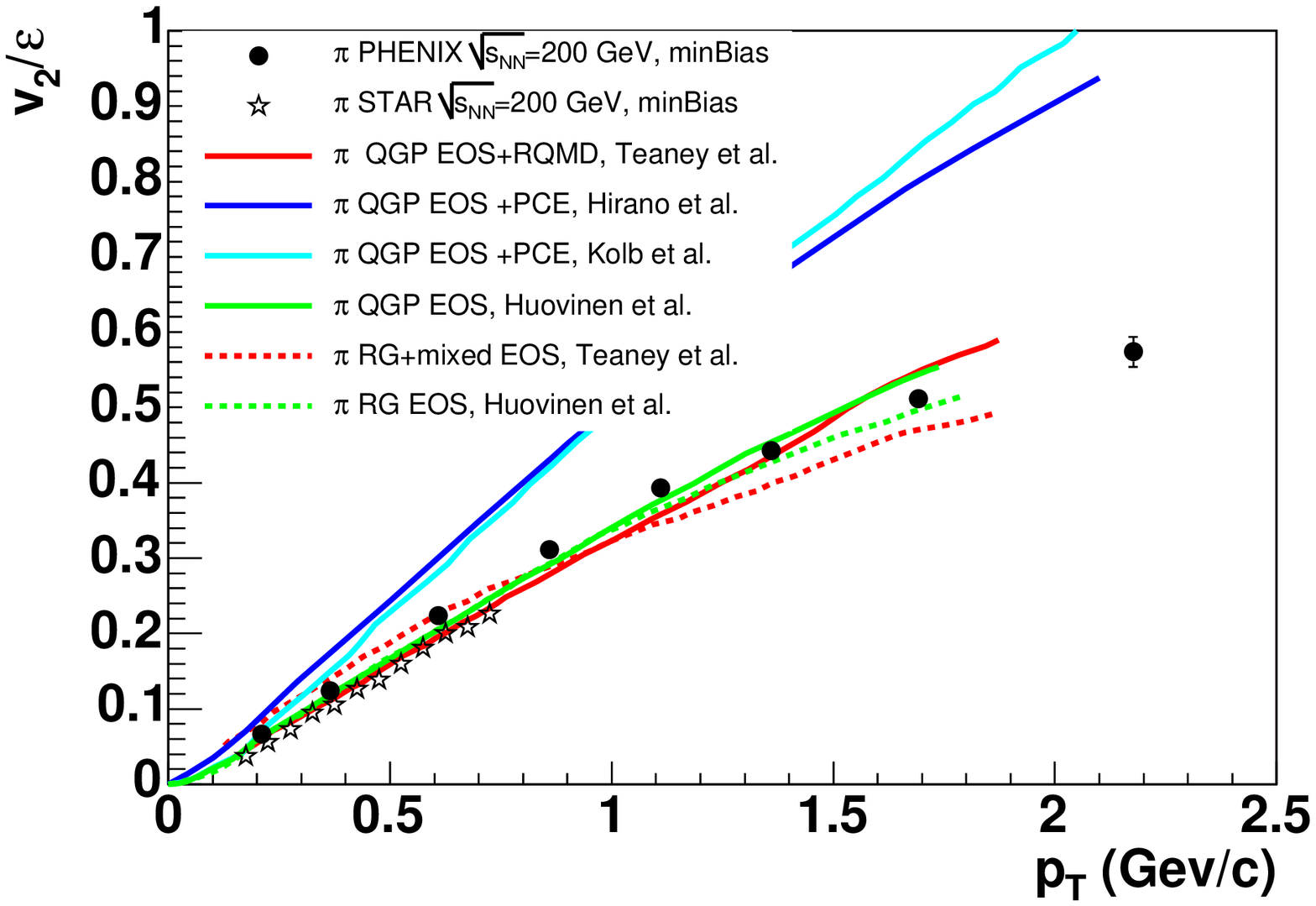} \\ [0.4cm] 
   \includegraphics[width=0.51\linewidth]{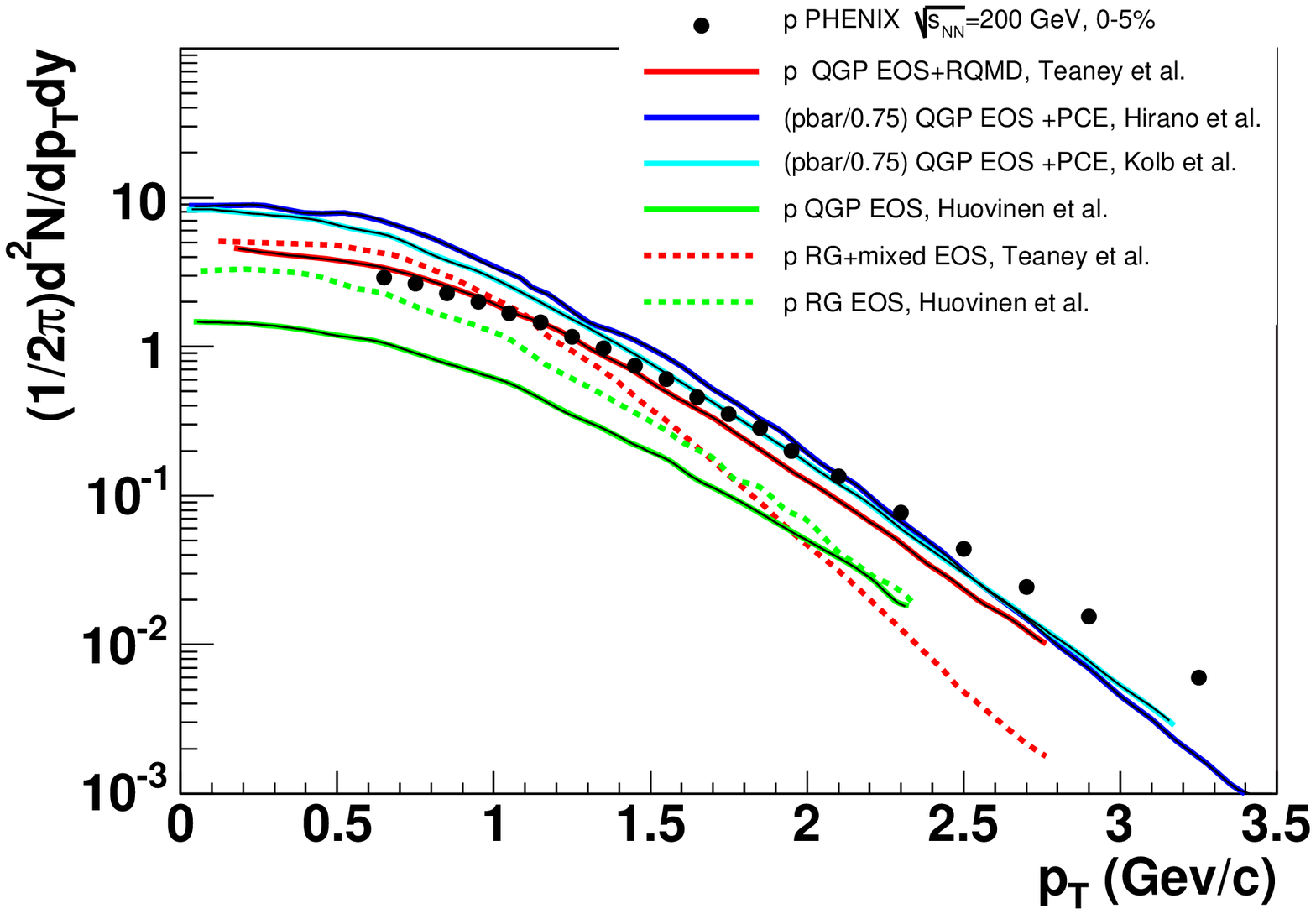} &
   \includegraphics[width=0.51\linewidth]{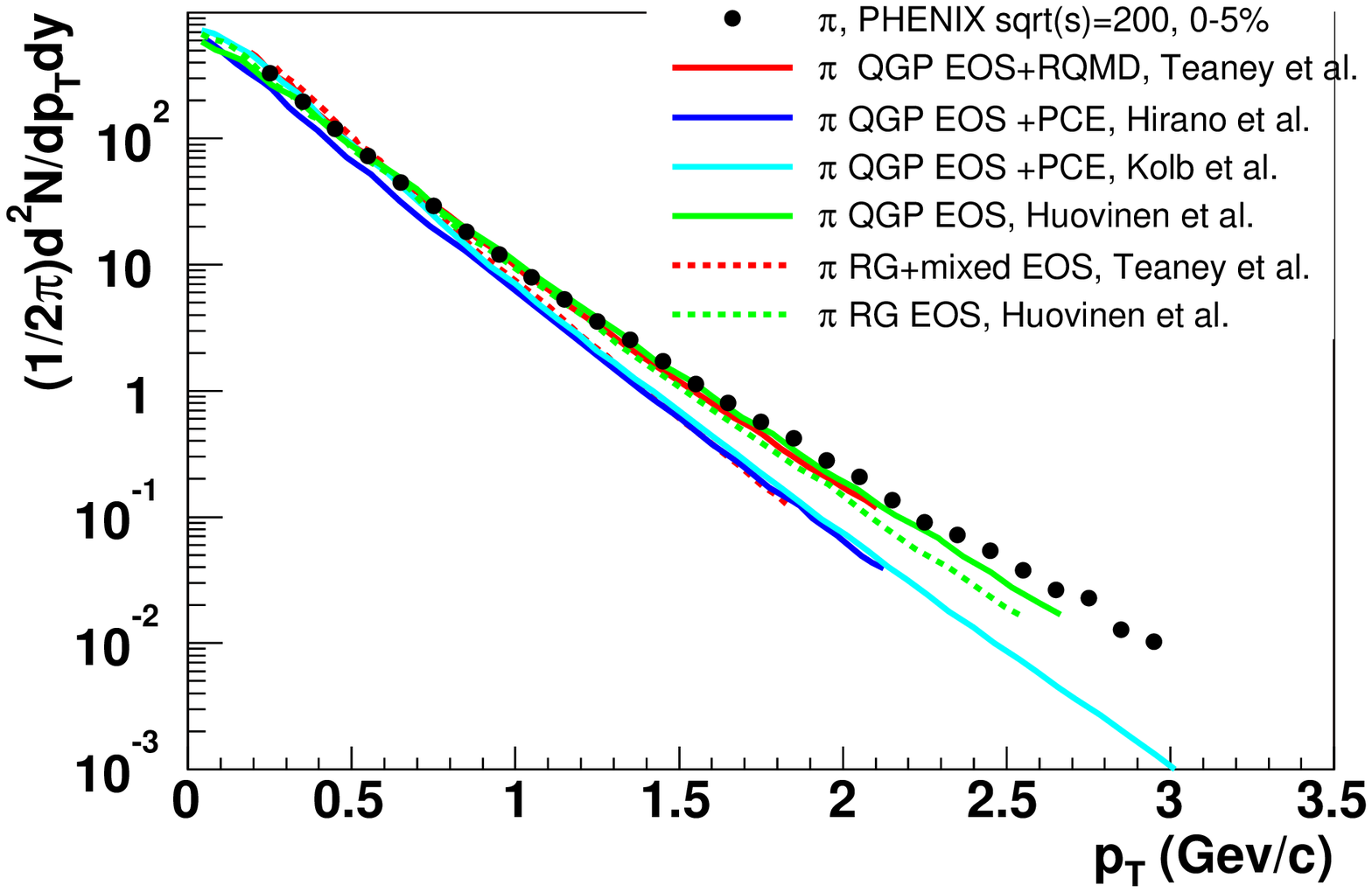} \\ [0.4cm]
\end{array}$
\caption{\label{fig:hydro} 
Top panels: proton and pion $1/\epsilon v_2(p_T)$ vs. $p_T$ for
minimum bias collisions at RHIC, compared with hydrodynamics calculations.
Bottom panels: proton and pion $p_T$ spectra for the most central 
(0-5\%) collisions at RHIC compared with the same hydrodynamics
calculations.
See text and \cite{phenix_wp} for references to the data and
hydrodynamics calculations. 
}
\end{center}
\end{figure}
%%%%%%%%%%%%%%%%%%%%%

Figure~\ref{fig:hydro} shows the comparison of a number of hydrodynamics
models \cite{teaney,huovinen,hirano,kolb}
with the data on pion and proton $v_2$ as a function of $p_T$ for 
minimum bias Au+Au collisions, and pion and proton $p_T$ spectra
in central collisions \cite{phenix_wp}. The calculations including
a phase transition from QGP to a hadronic phase are shown with solid
lines, while calculations with no pure QGP phase are drawn with
dashed lines. The four calculations that include a QGP phase all 
assume an ideal gas EOS for the QGP phase, a resonance gas for the
hadronic phase, and connect the two using a first-order phase
transition and a Maxwell construction. The calculations that do
not include a QGP phase either have a hadronic plus mixed phase
with the latent heat of the transition forced to infinity 
(Teaney)\cite{teaney}, 
or they use only a hadronic resonance gas (i.e. no mixed or QGP
phases in one of Huovinen's calculations) \cite{huovinen}. The 
calculations also differ in their
treatment of the final hadronic phase and freezeout. Hirano's
\cite{hirano}
is the only 3D hydrodynamics calculation; this and Kolb's \cite{kolb}
both allow for partial chemical equilibrium by chemically freezing
out earlier than the kinetic freezeout. This has been done in order
to reproduce the large proton yield measured at RHIC. 
In contrast, Huovinen\cite{huovinen} maintains full chemical 
equilibrium throughout the hadronic phase. Teaney\cite{teaney}
uses a hybrid model that couples the hadronic phase to RQMD to allow
hadrons to freeze out according to their scattering cross sections; 
incorporating this step allows for
chemical equilibrium to be broken in the hadronic phase. All four
models have assumed ideal hydrodynamics, i.e. zero viscosity and
zero mean free path.

From the top panels of Figure~\ref{fig:hydro} it is clear that 
the four calculations
that include a phase transition from the QGP phase to a hadronic
phase (solid lines) reproduce the low $p_T$ proton data better than 
the two hydro calculations without the QGP phase (dashed lines).
The presence of a first order phase transition softens the equation
of state, reducing the elliptic flow. At higher $p_T$ there is quite
some variation between the models. Part of this is due to the
modeling of the final hadronic stage. It is notable that Kolb's
and Hirano's calculations significantly overpredict $v_2$, and
they both have similar partial chemical equilibrium assumptions
in the late hadronic stage. Comparing to the transverse momentum
spectra, we see that all the models reproduce the pion spectra
below 1 GeV/c $p_T$. Calculations including a QGP phase do a better
job reproducing the proton spectra, presumably because of increased
transverse flow from the stronger early pressure gradients. The
calculations with partial chemical equilibrium during the hadronic
phase overpredict the proton spectra at low $p_T$.

All the models qualitatively reproduce the trends observed in the
data, but significant sensitivity to the assumptions in the models
is demonstrated. Thus is it important to reduce the model uncertainty
in the final state to extract quantitative information on the equation
of state during the reaction, including the possible softening of the
equation of state due to the presence of a mixed phase.
Nevertheless, we can use the calculations to extract limits for
the thermalization time\cite{phenix_wp}. All the hydrodynamical 
models require
quite short thermalization times, in the range of 0.6-1.0 fm/c,
to reproduce the magnitude of the observed elliptic flow. 

%\subsection{Heavy Quark Flow}

The general success of hydrodynamics models to explain the data
provides strong evidence for local thermalization of the system, 
or at least for equipartition of the momenta, in the
early stage of the collision. It is natural to ask whether the
heavy quarks flow along with the light ones, or whether their
large masses increase the time required for them to thermalize
so much that they cannot receive the same velocity boost from
the collective motion as the light quarks. 

Electrons are a useful
tool for the study of heavy quarks such as charm and bottom.
PHENIX has measured single electron $p_T$ spectra in Au+Au
collisions at $\sqrt{s_{NN}}$ = 130 GeV\cite{130charm} and
200 GeV\cite{charm}. The results are consistent with 
semileptonic charm decays in addition to the decays
of light mesons and photon conversions. The spectrum of
electrons of non-photonic origin can be measured by adding
a converter of well-calibrated thickness to determine the
photonic electron rate and subtracting these from the inclusive
electron spectrum. Alternatively, the measured $\pi^0$ and
$\eta$ spectra can be used to calculate the
photonic sources of electrons, including conversions of
photons from light
meson decays. The calculated background spectrum is then
subtracted from the measured electrons. The former technique
was used in \cite{charm}, and the latter to measure charm
production in p+p collisions. We use the same approach to
determine $v_2$ for open charm by measuring the distribution
of electrons with respect to the reaction plane. Though
electrons originating from decays of $D$ mesons have a
significant angular deviation from the original $D$ meson
directions, the extracted $v_2$ value remains well correlated
with the $v_2$ of the $D$ meson\cite{Greco}. Consequently,
PHENIX uses electrons and positrons to determine $v_2$ for
heavy quarks. We subtract $v_2$ of electrons from photon conversions
and Dalitz decays of light neutral mesons from the inclusive
electron $v_2$\cite{charmflow}.

%%%%%%%%%%%%%%%%%%%%%%%%%%%%%%%%%%%%%%%%%%%% Fig. 6
\begin{figure}
\begin{center}
\includegraphics[width=0.6\linewidth]{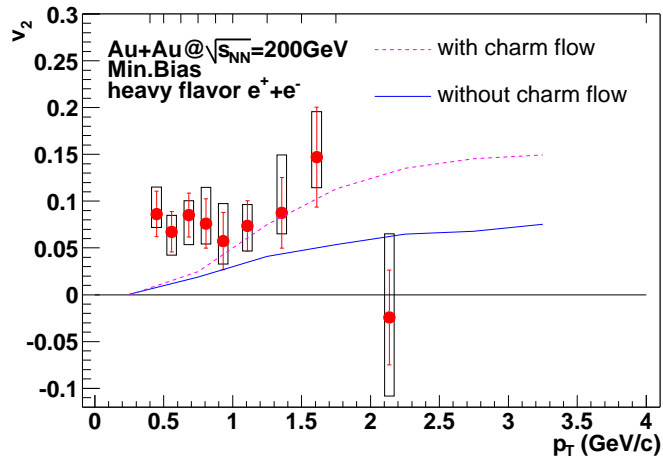}
\caption{\label{fig:charmflow} 
Elliptic flow ($v_2$) for electrons and positrons from heavy 
flavor decays, as a function of the electron or positron $p_T$,
in minimum bias $sqrt{s_{NN}}$ = 200 GeV Au+Au collisions. The
lines correspond to calculations with two different charm flow
scenarios\cite{Greco}. The solid line has no rescattering of the
initially produced charm quarks (i.e. no charm flow), while the
dashed line reflects the effect of complete thermalization (maximum
charm flow). 
}
\end{center}
\end{figure}
%%%%%%%%%%%%%%%%%%%%%

Figure~\ref{fig:charmflow} shows the heavy flavor $v_2$ in minimum
bias $\sqrt{s_{NN}}$ = 200 GeV Au+Au collisions. The statistical error
bars are propagated from the statistical uncertainties on the inclusive
electron $v_2$, while the bands show the 1$\sigma$ systematic uncertainty
of the heavy flavor $v_2$. The result allows calculation of the confidence
level for a non-zero $v_2$, yielding 90\% confifdence that the heavy
flavor electron $v_2$ is non-zero in the $p_T$ range from 1 - 1.75
GeV/c. The $v_2$ of electrons from decays of $D$ mesons have been
predicted assuming that the $D$ mesons are formed by charm quark
coalescence with thermal light quarks. The solid
line on Figure~\ref{fig:charmflow} assumes that the charm and 
anticharm quarks experience no reinteractions after they are formed
in initial state hard processes. The second scenario, shown by the
dashed line, assumes complete thermalization of the heavy quarks
with the transverse flow of the bulk matter. Though the large
statistical errors on this data sample do not allow exclusion of
either scenario, the points do suggest that heavy quark flow
may indeed be taking place.

\section{Probes of the Partonic State}

Given the various pieces of evidence pointing toward early
local thermalization of the system, it is of great interest to
use the short wavelength probes discussed in the Introduction
to measure the properties of the hot, dense matter 
$\approx$ 1 fm/c after the Au nuclei traverse one another.
Of paramount interest is how the the probes couple to the medium.
It is already well established that hard scattered partons traversing
the medium experience considerable energy loss\cite{ppg003}. But
the exact nature of the interaction of the parton with the medium
is not well studied experimentally. This question is particularly
interesting for partons of moderate $p_T$, between 5 and 10 GeV/c,
which may begin to hadronize in or near the medium.

%\subsection{Heavy Quark Energy Loss}

Having seen the tantalizing result that heavy quarks may indeed
 flow along with the bulk of light quarks, it is 
natural to wonder whether heavy quarks lose energy in the
medium as the light
quarks do. It has been predicted that the energy loss of charm
quarks should be less than that of light quarks, as the large 
charm quark mass mass decreases the phase space available for 
gluon radiation (i.e. "dead cone" effect)\cite{dima2}. The
dead cone should be even more significant for bottom quarks. More
detailed calculations include also simulation of the effects
of the charm $p_T$ spectral shape and contributions from $B$
meson decays; these predict that the nuclear modification factor
for electrons from charm decays should be 0.6-0.8 for electrons
with $p_T \approx$ 2.5 GeV/c \cite{magdalena}. This is a 
significant suppression, though less than that observed for light
quarks\cite{ppg003,ppg014}. Armesto and co-workers take into
account differences between the energy loss of quarks and gluons
traversing the medium, and predict even larger suppression of
electrons from heavy quarks\cite{armesto}. Moore and Teaney point
out that if the charm quarks demonstrate flow along with the
bulk of the medium, this is evidence for thermalization of charm,
and then the medium modifications of the charm spectrum should
be substantial\cite{mandt}.

%%%%%%%%%%%%%%%%%%%%%%%%%%%%%%%%%%%%%%%%%%%% Fig. 7
\begin{figure}
\begin{center}
\includegraphics[width=0.7\linewidth]{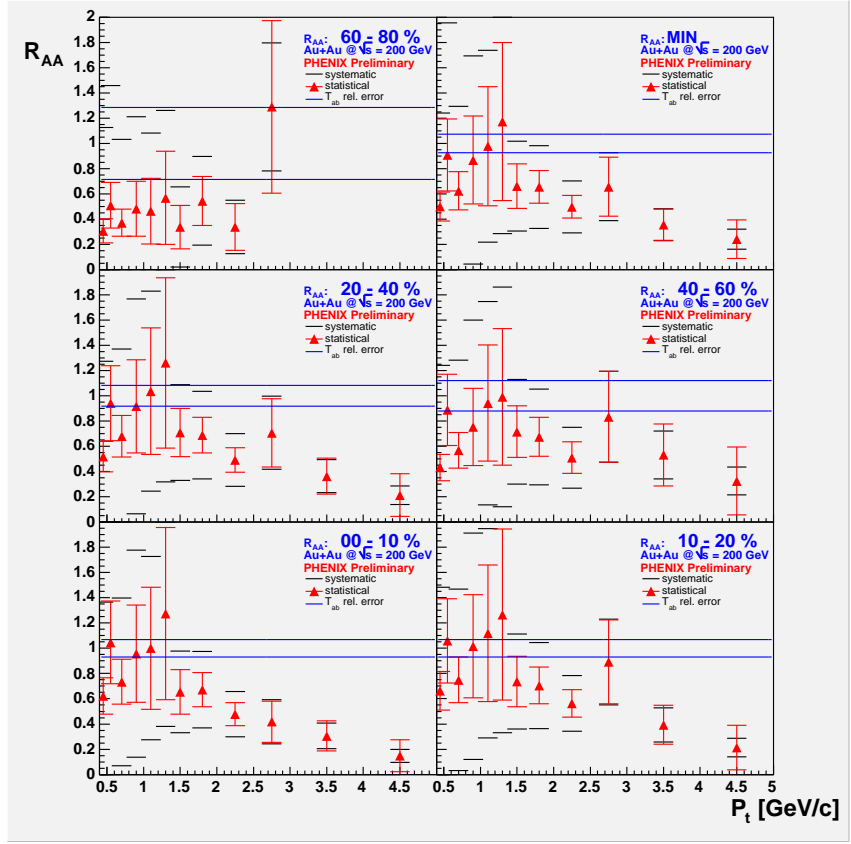}
\caption{\label{fig:charmeloss} 
Nuclear modification
factor, $R_{AA}$, for electrons and positrons from heavy quark
decays in $/sqrt{s_{NN}}$ = 200 GeV Au+Au collisions. The 
error bars show statistical errors, while the bands around
each point indicate systematic uncertainties. The bands
centered about $R_{AA}$ = 1.0 indicate the systematic
uncertainty in the number of binary nucleon collisions in
each centrality class.
}
\end{center}
\end{figure}
%%%%%%%%%%%%%%%%%%%%%%%%%%%%%%%%%%%%%%%%%%%%%%%%

Figure~\ref{fig:charmeloss} shows the nuclear modification
factor, $R_{AA}$ for electrons and positrons from heavy quark
decays in $/sqrt{s_{NN}}$ = 200 GeV Au+Au collisions, as a
function of the electron $p_T$. The figure shows minimum
bias collisions in addition to 5 centrality classes.  
Electrons from heavy quark decays are measured by subtracting
the $p_T$ spectrum of electrons from photonic sources, calculated
using the measured $\pi^0$ spectra, from the inclusive electron
spectrum. The same subtraction technique is used for Au+Au and
reference p+p spectra, but of course the input $\pi^0$ spectra
are different. In the most peripheral collisions, statistical
uncertainties limit the accessible $p_T$ range, but where there
are data, all the points are consistent with $R_{AA}$ = 1, i.e.
no suppression. In more central collisions, $R_{AA}$ falls well
below 1.0 for  electron $p_T \ge$ 2 GeV/c, providing clear
evidence for charm energy loss. At the highest $p_T$ of this
measurement, the electron $R_{AA}$ becomes nearly as small as that
for $\pi^0$. This is remarkable, as electrons above 3.5 or 4.0
GeV/c $p_T$ are expected to also include contributions from
$B$ meson decays, and $B$ mesons should experience less energy
loss than $D$ mesons. Though the data are consistent with predictions
from Armesto et al.\cite{armesto} for medium densities at the
extreme high end of those allowed by the observed light quark energy
loss, the lack of observable dilution from $B$ meson decays is
intriguing.

%\subsection{Jet Modification}

Hard scattered partons are formed 
in the initial nucleon collisions,
and traverse the hot, dense medium.
These partons experience significant energy loss 
via medium-induced gluon radiation. The next question 
is - where does the radiated energy go?
Though radiated gluons are nearly collinear with the
parent parton, they are rather soft and may interact
further with the gluon field in the medium. 
%Theoretical descriptions of jet quenching in nuclear collisions
%often rely upon the notion of collinear factorization, 
%treating the intial hard scattering, induced gluon radiation, and
%fragmentation into a jet of hadrons as processes well separated 
%in time. 
Though it is often assumed that fragmentation
takes place outside the medium, the formation time for 2-4
GeV/c $p_T$ hadrons is not very long, particularly
for baryons. Thus fragmentation may begin in or near the
medium and involve
not only the quarks and antiquarks from gluons radiated as 
fragmentation begins, but also comoving gluons from 
medium-induced radiation, and indeed recombination of jet partons 
with partons from the medium itself\cite{Hwa}. 

PHENIX probes these processes using correlations of high transverse
momentum particles, in the $p_T$ range from 1 - 4 GeV/c. We 
select Au+Au collisions containing hard scattering events by
requiring detection of a hadron with 2.5 $\le p_T \le$ 
4 GeV/c. We then construct correlation functions between these
trigger particles and associated particles with 1.0 $\le p_T
\le$ 2.5 GeV/c, as a function of their azimuthal angle 
difference\cite{ajit,ppg032}. 
%As the PHENIX acceptance at central rapidity is non-uniform in azimuth, 
We correct for the non-uniform PHENIX azimuthal 
acceptance by constructing an area normalized correlation
function utilizing pairs from mixed events. To extract the
yield of jet-induced pairs, we analyze the correlation function
assuming that each hadron can be attributed either to a jet
fragmentation source or to the underlying event. The
underlying event has an azimuthal correlation arising from
the fact that the single particle
distributions respect the reaction plane of
the event. This correlation is removed by modulating the
background pair distribution with the inclusive $v_2$
value measured for the trigger and associated particle
$p_T$ range. Thus the correlation function is decomposed via
$$ C(\Delta\phi) = 
b_0(1+2(v_2^T v_2^A)cos(2\Delta\phi)) + J(\Delta\phi)$$
The average level of the background, $b_0$ is fixed by
assuming that the jet fragmentation yield of particle pairs
is zero for at least one value of $\Delta\phi$. We refer
to this as the ZYAM (zero yield at minimum) assumption. We
have verified the validity of this assumption by independently
estimating the $b_0$ values using the T*A combinatorial pair
rate. Once the background level is determined, we extract 
conditional yields for particle pairs from jets.

%%%%%%%%%%%%%%%%%%%%%%%%%%%%%%%%%%%%%%%%%%%% Fig. 8
\begin{figure}
\begin{center}
\includegraphics[width=0.7\linewidth]{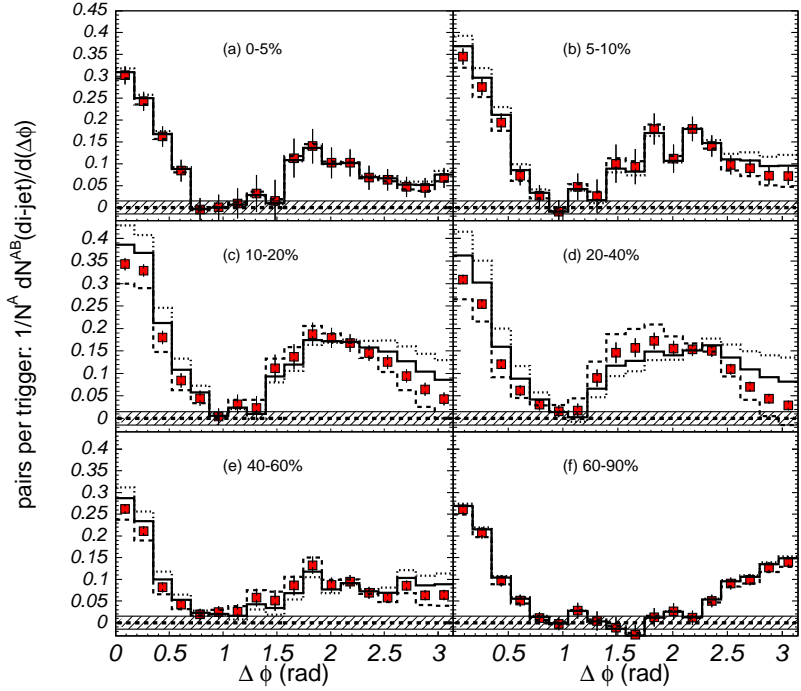}
\caption{\label{fig:jetmod} 
Jet fragment pair relative azimuthal angle distributions for different
centralities in $\sqrt{s_{NN}}$ = 200 GeV Au+Au collisions\cite{ppg032}. 
The yields are normalized per trigger particle. The shaded bands indicate
the systematic error associated with the determination of the 
relative angle where the jet pair yield is zero. The dashed(solid)
lines are the distributions that would result from increasing
(decreasing) $\langle v_2^T v_2^A \rangle$ by one unit of the systematic
error; the dotted curve would result from decreasing by two units.
}
\end{center}
\end{figure}
%%%%%%%%%%%%%%%%%%%%%%%%%%%%%%%%%%%%%%%%%%%%%%%%

The conditional yields of jet partners per trigger particle
are shown in Figure~\ref{fig:jetmod}\cite{ppg032}. For the most peripheral
event sample, the jet associated yield distribution has the
same shape as jet pairs in p+p collisions: a well-defined near
side peak around $\Delta\phi$ = 0 and a somewhat wider away side
peak around $\Delta\phi = \pi$. For more central event samples,
the shape of the near-side peak is essentially unchanged while
the associated yield in the near-side peak increases, indicating
some change in the fragmentation process. A much more
dramatic change is visible in the away side peak, which is
considerably broader in all the centrality classes more central
than 60\%. In mid-central and central collisons there is a local
minimum at $\Delta\phi = \pi$. Though the existence of these
local minima $per se$ is not significant once we take the 
systematic errors on $\langle v_2^T v_2^A \rangle$ into account,
it is clear that the away side peak in more central collisions
has a very different shape than in peripheral collisions. 

Convoluting the jet fragments' angles with respect to their
parent partons and the acoplanarity between the two partons
would yield a Gaussian-like shape in $\Delta\phi$, possibly
broadened through jet quenching. The observed shapes in the
away side peaks cannot result from such a convolution.
The away side peaks are suggestive of recent theoretical 
predictions of dense medium effects on fragment distributions.
These include combination of jet partons with medium partons 
accelerated
by a density wave in the shocked medium\cite{jorge,horst,berndt},
and jet asymmetries caused by the hydrodynamic flow of the 
underlying event\cite{armesto2}. The broadening of the away
side jet implies that integration of the away side peak in a
narrow angular range around $\Delta\phi = \pi$ yields fewer
associated partners in central collisions than in peripheral
Au+Au or in d+Au collisions. However, integrating over the
entire broadened peak recovers the jet partners in the range
1.0 GeV/c $\le p_T^A \le$ 2.5 GeV/c used in this
analysis.

\section{Summary}

The PHENIX collaboration has measured a large
number of observables in Au+Au, d+Au and p+p collisions at RHIC.
We have shown that direct photons and hard processes are calculable
with perturbative QCD, yielding a calibrated source of short
wavelength probes of the medium formed in nuclear collisions.
%We observe nuclear shadowing at RHIC. There is no room
%for saturation of the gluon distribution at mid-rapidity, but
%a color glass condensate could be present at forward rapidities.
Anomalous behavior of baryon yields is seen in d+Au 
collisions, suggesting that baryon formation is already subject
to the influence of the nearby nucleus in that case.

The elliptic flow trends
support the picture of rapid
thermalization in Au+Au collisions. Furthermore, there are 
first indications that heavy quarks may participate in the
collective flow with the light quarks.
%The theoretical community has some homework to do, however, 
%to determine just how soft the equation of state can be. 
We have seen that heavy quarks 
lose energy in the hot dense medium, as may be expected if
they approach local thermal equilibrium.
Jet fragmentation is modified in central and 
semi-central Au+Au collisions. The data suggest that the lost
energy may excite the medium and modify the formation of jet
hadrons. The formation of moderate
$p_T$ baryons (from approximately 2-5 GeV/c) is expected to
take place in, or near, the medium in Au+Au and d+Au collisions.
Indeed the data show enhanced baryon production probability 
in both cases. This suggests that recombination of partons
from the fragmenting jet and from the medium play an
important role in hadronization.

Quantitative determination of the opacity of and collision
frequency in the medium, along with its color Debye screening
length awaits completion of the analysis of the high statistics
run4 data. PHENIX has a billion events currently being analyzed.

\section{References}

\def\Journal#1#2#3#4{{#1}{\bf #2}, #3 (#4)}
\def\IJMPA{{Int. J. Mod. Phys.}~{\bf A}}
\def\JPG{{J. Phys}~{\bf G}}
\def\NCA{Nuovo Cimento\ }
\def\NIM{Nucl. Instrum. Methods\ }
\def\NIMA{{Nucl. Instrum. Methods\ }~{\bf A}}
\def\NPA{{Nucl. Phys.}~{\bf A}}
\def\NPB{{Nucl. Phys.}~{\bf B}}
\def\PLB{{Phys. Lett.}~{\bf B}}
\def\PLC{Phys. Repts.\ }
\def\PR{Phys. Rev.\ }
\def\PRL{Phys. Rev. Lett.\ }
\def\PRD{Phys. Rev. D\ }
\def\PRC{Phys. Rev. C\ }
\def\RMP{Rev. Mod. Phys.\ }
\def\SPJ{Sov. Phys. JETP\ }
\def\SJNP{Sov. J. Nucl. Phys.\ }
\def\ZPC{{Z. Phys.}~{\bf C}}

\end{document}